\documentclass[a4paper]{IEEEtran}
\usepackage{color}
\usepackage{leftidx}
\usepackage{cite}
\usepackage{microtype}
\usepackage{cleveref}
\usepackage{graphicx,amssymb,amstext,amsmath}
\usepackage[ruled,vlined,lined,ruled,linesnumbered]{algorithm2e}
\usepackage{float}

\begin{document}

\title{Reusing Wireless Power Transfer for Backscatter-assisted Relaying in WPCNs}
\author{Yuan~Zheng,~Suzhi~Bi, Xiaohui~Lin, and Hui~Wang\\
\thanks{This work has been presented in part in the International Conference on Machine Learning and Intelligent Communications (MLICOM), Hangzhou, China, July 2018 \cite{2018:Zheng}.}
\thanks{The authors are with the College of Electronics and Information Engineering, Shenzhen University, Shenzhen, Guangdong, China 518060. E-mail:~\{zhyu,~bsz,~xhlin,~wanghsz\}@szu.edu.cn }
\vspace{-2ex}}

\maketitle

\begin{abstract}
   User cooperation is an effective technique to tackle the severe near-far user unfairness problem in wireless powered communication networks (WPCNs). In this paper, we consider a WPCN where two collaborating wireless devices (WDs) first harvest wireless energy from a hybrid access point (HAP) and then transmit their information to the HAP. The WD with the stronger WD-to-HAP channel helps relay the message of the other weaker user. In particular, we exploit the use of ambient backscatter communication during the wireless energy transfer phase, where the weaker user backscatters the received energy signal to transmit its information to the relay user in a passive manner. By doing so, the relay user can reuse the energy signal for simultaneous energy harvesting and information decoding (e.g., using an energy detector). Compared to active information transmission in conventional WPCNs, the proposed method effectively saves the energy and time consumed by the weaker user on information transmission during cooperation. With the proposed backscatter-assisted relaying scheme, we jointly optimize the time and power allocations on wireless energy and information transmissions to maximize the common throughput. Specifically, we derive the semi-closed-form expressions of the optimal solution and propose a low-complexity optimal algorithm to solve the joint optimization problem. By comparing with some representative benchmark methods, we simulate under extensive network setups and demonstrate that the proposed cooperation method effectively improves the throughput performance in WPCNs.
\end{abstract}
\vspace{-2ex}

\IEEEpeerreviewmaketitle

\section{Introduction}
The limited battery lifetime is a crucial factor  affecting the performance of wireless communications. Wireless devices (WDs) need to replace/recharge battery when the energy is exhausted, which leads to frequent interruption to normal communication process and severe degradation of the quality of communication service. Alternatively, thanks to the recent advance of radio frequency (RF) based wireless energy transfer (WET) technology, the WDs can continuously harvest energy without interrupting their normal operation. The newly emerged wireless powered communication network (WPCN) integrates WET into conventional wireless communication system \cite{2014:Bi,2015:Lu,2014:Ju1,2016:Bi2,2017:Bi,2018:Bi,2019:Bi}, which has shown its advantages in lowering the operating cost and improving the robustness of communication service in low power applications, such as sensing devices in internet of things (IoT) networks. There have been extensive studies on the design and optimization in WPCN. For instance, \cite{2014:Ju1} presented a harvest-then-transmit strategy in WPCN, where WDs first harvest RF energy from a single antenna hybrid access point (HAP) in the downlink (DL), and then use the harvested energy to transmit information to the HAP in a time-division-multiple-access (TDMA) manner in the uplink (UL). Besides, \cite{2014:Ju1} revealed an inherent doubly near-far problem in WPCN, where the near user from the HAP achieves much higher transmission rate than the farther user as it harvests more energy from and consumes less energy to transmit information to the HAP. To improve the user fairness, \cite{2014:Ju2,2017:MM,2018:SP,2017:Yuan} have proposed several different user cooperation methods. For example, a two-user cooperation WPCN was presented in \cite{2014:Ju2}, where the near user with more abundant energy helps relay the far user's information to the HAP. Besides, \cite{2017:MM} allowed two cooperating users to form a distributed virtual antenna array and transmit jointly to the information access point. \cite{2018:SP} considered optimal transceiver design and relay selection for simultaneous wireless information and power transfer (SWIPT) in a two-hop cooperative network with energy harvesting constraints at the receiver. Further, the authors in \cite{2017:Yuan} proposed a cluster-based user cooperation method, where one of a cluster of users is designated as the cluster head to relay the other users' information. To supplement the higher energy consumption of the cluster head, the multi-antenna HAP applies the energy beamforming technique \cite{2014:Bi} to achieve directional energy transfer.

A major concern in the design of user cooperation in WPCN is the time and energy consumption on exchanging individual information among the collaborating users. Recently,  ambient backscatter (AB) communication technology has emerged as a promising method to reduce the cooperation overhead \cite{2014:AAA,2014:BAJ,2019:Cao,2014:PD}. Specifically, with AB communication, a WD can backscatter the RF signal (e.g., WiFi and cellular signals) to transmit its information to another WD in a passive manner \cite{2018:ND}, thus saving the device battery on generating and transmitting carrier signals as in conventional active information transmissions. Several recent works have studied signal detection methods \cite{2013:Vli} and communication circuit design \cite{2016:Gw} to improve the throughput of AB communication.
In practice, the performance of AB communication has been evaluated in various wireless scenarios, where \cite{2017:DDG} showed that AB communication achieves high transmission rates over relatively short distances, e.g., less than $10$ meters. \cite{2015:DKM} developed a BackFi backscatter system that improves communication rates up to 5Mbps within 1m and 1Mbps within 5m in the backscatter communication link using ambient WiFi signals, \cite{2018:JA} employed the high-order ($M$-PSK) modulation for AB communication and devised the corresponding maximum likelihood detector, \cite{2019:Q} analyzed the achievable rate and capacity for AB communication with the instantaneous channel state information (CSI). In addition, \cite{2017:KK} presented a network architecture for a large-scale backscatter communication network, modeled and analyzed the communication performance using stochastic geometry.

\begin{figure}
  \centering
   \begin{center}
      \includegraphics[width=0.5\textwidth]{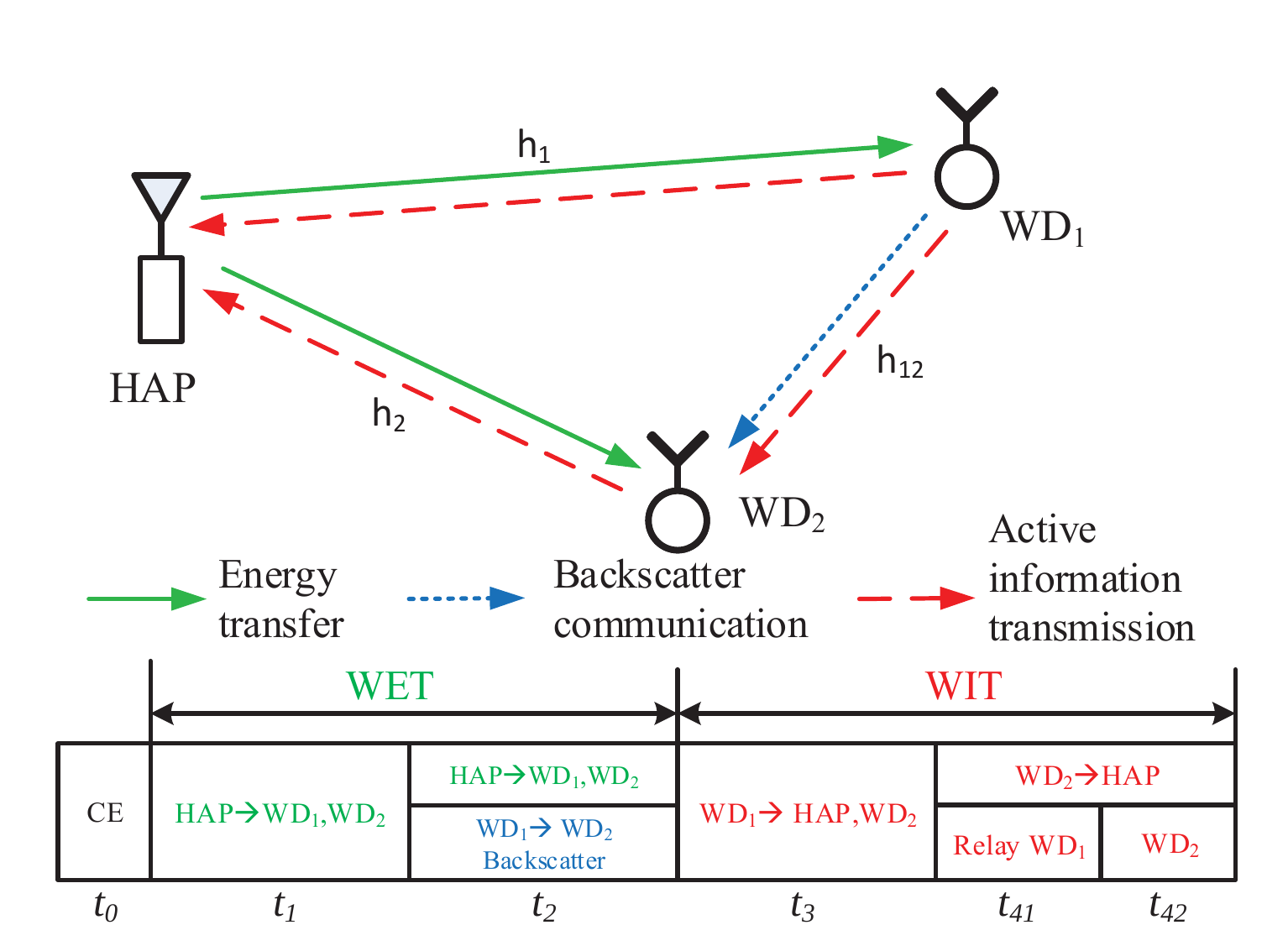}
   \end{center}
  \caption{The network structure and transmission strategy of the proposed cooperation scheme. }
  \label{Fig.1}
\end{figure}

The integration of AB communication technique in modern communication network leads to many new technological innovations and networking paradigms. However, a major performance limitation is the time-varying ambient RF signal source, whose randomness in both strength and time availability renders AB communication performance uncontrollable. The combination of WET technology and AB communication effectively mitigates such problem, where the fully controllable energy signal is used as the carrier of  AB communication \cite{2015:GCY,2017:DTH,2018:DG,2017:SD}. For instance, \cite{2015:GCY} optimized the energy beamforming from a multi-antenna energy transmitter to multiple energy receivers with limited channel estimations at destined receivers in a backscatter communication system. \cite{2017:DTH} and \cite{2018:DG} introduced AB communication into RF-powered cognitive radio networks, and showed the improved throughput performance of the secondary system. Further, \cite{2017:SD} investigated a hybrid wireless powered backscatter communication scheme in heterogeneous wireless networks. Overall, the combination of WPT and AB communications provides more robust and energy-conserving communication service in low-power applications.

Recently, several works have also examined the use of AB communication for cooperative transmissions in WPCN \cite{2018:GS,2018:DD,2019:LB}. For instance, a backscatter relay communication system powered by an energy beacon station was first studied in \cite{2018:GS}, where each backscatter radio harvests energy to sustain battery-less transmissions, while the other radios serve as relays to realize cooperative transmission. \cite{2018:DD} proposed a relay selection scheme for backscatter communications which enables the out-of-coverage device to communicate with the HAP via backscatter relay devices, in which the HAP adopts energy beamforming to power the backscatter devices to carry out their operations.
 \cite{2019:LB} presented two user cooperation schemes in a WPCN with backscatter communication, where one device operates in backscatter mode and the other device operates in harvest-then-transmit mode. The authors considered two cases in which either one of the two devices serves as the relay node for the other device in forwarding information to the AP to improve the overall throughput performance. However, most of the existing works that adopt AB communication for cooperation consider a collaborating device transmitting information in either active RF communication mode or passive backscatter communication mode. In practice, however, a device can harvest energy and receive information backscattered from the other device simultaneously during the wireless power transfer stage. Meanwhile, the harvested energy can be used to transmit information actively in later stage. Therefore, it is promising to implement cooperative transmissions in a WPCN by allowing a device to transmit both in active and passive communication. In this case, a joint design of system resource allocation on both active and passive communications is needed to achieve the maximum energy and communication efficiency. However, to the best of our knowledge, this important research topic is currently lacking of concrete study.

In this paper, we consider realizing efficient user cooperation in WPCN using both active RF communication and AB-assisted passive communication. In this system, WD$_1$ can be either in the active communication mode or the backscatter communication mode to transmit information to WD$_2$. As shown in Fig.~\ref{Fig.1}, we consider that an HAP broadcasts wireless energy to two WDs in the downlink and receives information transmission from the WDs in the uplink. Specifically, during the WET stage ($t_2$ time slot), the weaker user (WD$_1$) backscatters the received energy signal to transmit its information to the relay user (WD$_2$) in a passive manner. Meanwhile, the relay user can reuse the energy signal for simultaneous energy harvesting and information decoding using a non-coherent information decoder, e.g., energy detector. Such signal reuse effectively reduces the collaborating overhead compared to when conventional active information transmission is used.

The detailed contributions of this paper are summarized as follows:
\begin{itemize}
  \item The proposed user cooperation scheme exploits the use of AB communication during the WET stage, which enables the relay user to harvest energy from the HAP and receive the other user's information simultaneously. Compared to existing cooperation scheme without backscatter communication, the considered backscatter-assisted cooperation method reduces the collaborating overhead (transmission time and energy consumption) in the WPCN, and thus has the potential to improve the overall communication performance.
  \item With the considered AB-assisted cooperation scheme, we first analyze the achievable data rates of the two users. Then, we jointly optimize the system time and power allocations on wireless energy and information transmissions to maximize the common throughput, which is an important metric of user fairness in WPCN. We derive the semi-closed-form expressions of the optimal solution and propose an efficient algorithm to solve the optimization problem.
  \item We simulate under extensive network setups to evaluate the performance of the proposed backscatter-assisted cooperation method. By comparing with conventional user cooperation method based on active communication, we show that the proposed passive cooperation can effectively enhance the throughput performance of energy-constrained devices in WPCN, especially when the weaker user is unable to harvest sufficient energy for efficient active information transmission.
\end{itemize}

The rest of the paper is organized as follows: In Section II, we present the system model of the proposed backscatter-assisted relaying in WPCN. We formulate the max-min throughput optimization problem in Section III and propose an efficient algorithm to solve it in Section IV. In Section V, we perform simulations to evaluate the performance of the proposed cooperation method. Finally, Section VI concludes this paper.

\section{System Model}
\subsection{Channel Model}
As shown in Fig.~\ref{Fig.1}. we consider a WPCN where the HAP broadcasts RF energy to the two WDs in the DL and receives the WDs' information in the UL. The HAP and the two WDs are assumed to be equipped with one antenna each. We assume that all devices operate over the same frequency band. For simplicity of expression, it is assumed that the channel reciprocity holds for the communication links.  We denote $\alpha_i$ and $h_i=|\alpha_i|^2,i=1,2,$ as the channel coefficient and the channel power gain between the HAP and WD$_i$. Besides, the channel coefficient between WD$_1$ and WD$_2$ is $\alpha_{12}$ and the corresponding channel power gain is $h_{12}=|\alpha_{12}|^2$. Without loss of generality, we assume that WD$_2$ is closer to the HAP and has a better channel condition, such that it helps relay WD$_1$'s information to the HAP.

\begin{figure}
  \centering
   \begin{center}
      \includegraphics[width=0.4\textwidth]{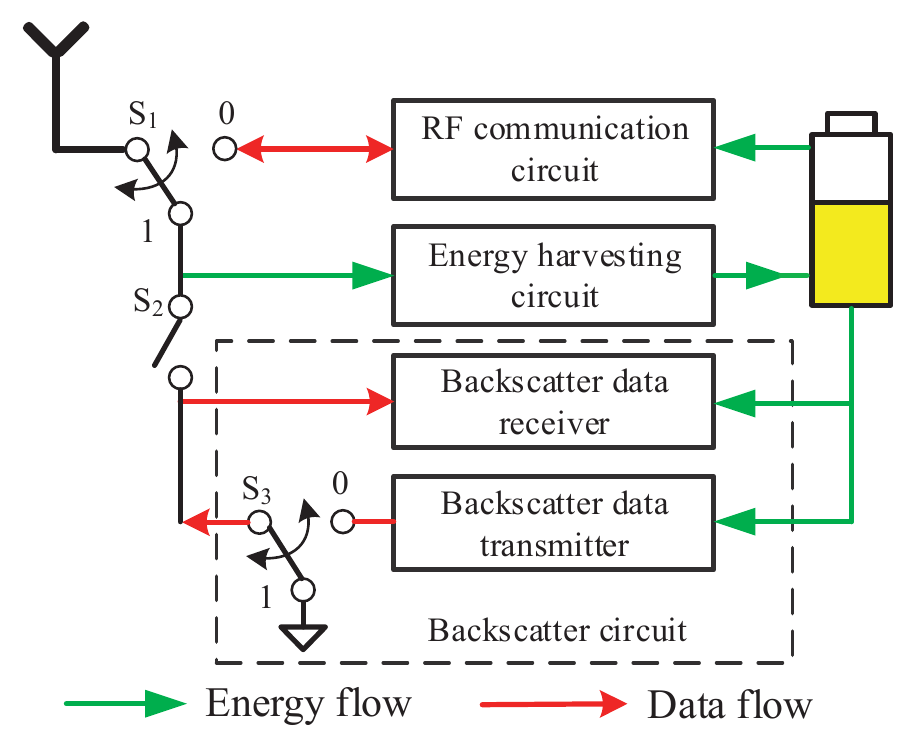}
   \end{center}
  \caption{Circuit block diagram of backscatter wireless user. }
  \label{Fig.2}
\end{figure}

The two users can perform information transmissions in two modes: active RF communication mode and passive backscatter communication  mode. We illustrate the circuit block diagram of two WDs in Fig.~\ref{Fig.2}. The two users can switch flexibly among the following three operating modes with the two switches $S_1$ and $S_2$.
\begin{enumerate}
  \item \emph{RF Communication Mode} ($S_1=0$): the active communication mode is activated when the RF communication circuit connects to the antenna. In this case, the WDs  apply traditional RF wireless communication techniques to transmit and receive information, e.g., using QAM encoder and coherent detector. The energy consumption of active transmission is powered by an on-chip rechargeable battery.
  \item \emph{Energy-harvesting Mode} ($S_1=1$ and $S_2$ is open): in this mode, the antenna is connected to the energy harvesting circuit, such that the received RF signal is converted into direct current energy and stored in a rechargeable battery, which supplies the power consumptions of the other circuits.
  \item \emph{Backscatter Mode} ($S_1=1$ and $S_2$ is closed): when the passive communication mode is used, energy harvesting and backscatter communication circuits are both connected to the antenna. Further, when setting the switch $S_3=1$, the circuit operates in the reflecting state to transmit ``1". Otherwise, when $S_3=0$, the circuit switches to the absorbing state and ``0" is transmitted.  Accordingly, the backscatter receiver decodes the 1-bit information using a non-coherent detection method, e.g., energy detector\cite{2012:MAK}. Notice that the energy harvesting circuit can harvest a small amount of energy during the backscatter mode especially when transmitting ``0". The harvested energy is sufficient to power the backscatter circuit, thus we neglect the energy consumption when performing backscatter communication (such as in \cite{2017:DTH}).
\end{enumerate}

\subsection{Protocol Description}
The time allocation of the proposed backscatter-assisted relaying is shown in Fig.~\ref{Fig.1}. Initially, channel estimation (CE) occupies the first time block of length $t_0$, from which the HAP (or a central control point) has the knowledge of channel coefficients $\{\alpha_1,\alpha_2,\alpha_{12}\}$, e.g., via channel sounding. Subsequently, the backscatter-assisted relaying communication consists of four operation phases. In the first phase, the HAP transfers wireless energy to the WDs in the DL for $t_1$ amount of time. In the second phase, WD$_1$ backscatters the received energy signal to transmit its information to WD$_2$ for $t_2$ amount of time. Notice that WD$_2$ can decode the backscattered information from the WD$_1$ and simultaneously harvest wireless power transfer from the HAP, which will be detailed in Section III. We assume that the HAP is only equipped with conventional active RF communication circuit such that it does not decode the reflected signal from WD$_1$. The case that the HAP also decodes from the reflected signal will be investigated in future study.

In the third phase of duration $t_3$, WD$_1$ uses the harvested energy to transmit its information to WD$_2$ in conventional active communication mode. Note that RF transmission of WD$_1$ can be overheard by the HAP during this phase. In the last phase of duration $t_4$, the WD$_2$ transmits information to the HAP. In particular, $t_4$ is divided into two parts.  In the first part of duration $t_{41}$, WD$_2$ acts as a relay to transmit WD$_1$'s information to the HAP. In the second part of duration $t_{42}$, WD$_2$ conveys its own message to the HAP, where $t_4=t_{41}+t_{42}$. Accordingly, the total time constraint is
\begin{equation}
\label{t}
\small
t_0+t_1+t_2+t_3+t_{41}+t_{42}\leq T.
\end{equation}

Without loss of generality, it is assumed that $t_0$ is a fixed parameter. In the following section, we derive the optimal throughput performance of the considered backscatter-assisted cooperation in WPCN.

\section{Throughput Performance Analysis}
\subsection{Phase \uppercase\expandafter{\romannumeral1}: Energy Transfer}
In the first stage of length $t_1$, the HAP transfers wireless energy to WD$_1$ and WD$_2$ with fixed transmit power $P_1$. We denote $x_1(t)$ as the baseband equivalent energy signal transmitted from the HAP, which is a pseudo-random sequence with $E[|x_1(t)|^2] =1$ \cite{2014:Bi}. Then, the two WDs receive
\begin{equation}
y_i^{(1)}(t) = \alpha_i \sqrt{P_1} x_1(t) + n_i^{(1)}(t), i=1,2,
\end{equation}
where $n_i(t)$ denotes the receiver noise power. It is assumed that the energy received from the receiver noise is negligible, where WD$_1$ and WD$_2$ harvest the following amount of energy in the first phase \cite{2016:Bii}
\begin{equation}\label{energy}
  E_1^{(1)}={\eta}{t_1}{P_1}{h_1},\  E_2^{(1)}={\eta}{t_1}{P_1}{h_2}.
\end{equation}
Here, $0\textless \eta \textless 1$ denotes the fixed energy harvesting efficiency.\footnote{Although a single energy harvesting circuit exhibits non-linear energy harvesting property due to the saturation effect of circuit, it is shown that the non-linear effect can be effectively rectified by using multiple energy harvesting circuits concatenated in parallel, resulting in a sufficiently large linear conversion region in practice \cite{2019:Kang,2019:Ma}. }

\subsection{Phase \uppercase\expandafter{\romannumeral2}: Backscatter Information Transmission}

In the backscattering phase, WD$_1$ backscatters the received energy signal to transmit its information to WD$_2$ for $t_2$ amount of time. We denote the baseband equivalent pseudo-random energy signal transmitted by the HAP as $x_2(t)$ with $E[|x_2(t)|^2] =1$.  We assume that the backscattering transmission rate is $R_b$ bits/second,  which is a fixed parameter determined by the backscatter circuit, thus it takes $1/R_b$ second to transmit one bit information. Specially, when a symbol ``0'' is transmitted by WD$_1$, the  WD$_2$ receives only the energy signal from the HAP, which is expressed  as
\begin{equation}
\label{202}
y_{2,0}^{(2)}(t) = \alpha_2 \sqrt{P_1} x_2(t) + n_2^{(2)}(t).
\end{equation}
Otherwise, when a symbol ``1'' is transmitted, WD$_2$ receives the energy signal and WD$_1$'s reflected signal, i.e.,
\begin{equation}
\label{212}
y_{2,1}^{(2)}(t) =\alpha_2 \sqrt{P_1}x_2(t) +  \mu\alpha_1\alpha_{12}\sqrt{P_1}x_2(t) + n_2^{(2)}(t),
\end{equation}
where $n_2^{(2)}(t)$ is the receiver noise at WD$_2$ with power $N_0$, and $\mu$ denotes the complex signal attenuation parameter of the reflection at WD$_1$ with $|\mu|\le1$.

\begin{figure}
  \centering
   \begin{center}
      \includegraphics[width=0.5\textwidth]{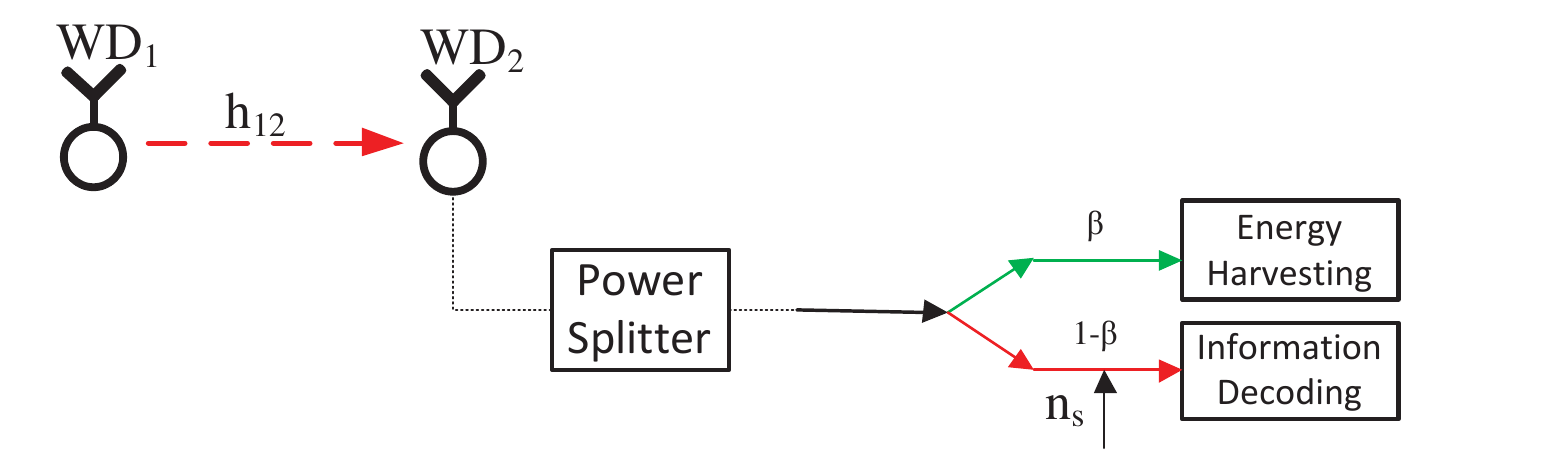}
   \end{center}
  \caption{The power splitting model in the backscattering phase. }
  \label{Fig.3}
\end{figure}

We consider implementing a power splitting receiver at WD$_2$ in Fig.~\ref{Fig.3}, where it can split the received RF signal into two parts. Specifically, $\beta$ of the signal power is harvested and stored in the battery, and the rest $(1-\beta)$ of the signal power is used for information decoding (ID), where $\beta\in[0,1]$ is the splitting factor. For convenience, we assume that $\beta$ is a constant in the following sections, and the impact of $\beta$ to the overall system performance will be investigated numerically in simulation. The information decoding circuit introduces an additional independent noise $n_s(t)$ with power $N_s$ \cite{2013:Zhou}. Thus, the energy and information signals at the WD$_2$ are

\begin{equation}
y_{2,E}^{(2)}(t)=\sqrt{\beta}y_{2}^{(2)}(t),
\end{equation}
\begin{equation}
y_{2,I}^{(2)}(t)=\sqrt{1-\beta}y_{2}^{(2)}(t)+n_s(t),
\end{equation}
where $y_2^{(2)}(t) = y_{2,0}^{(2)}(t)$ when transmitting ``0" and  $y_2^{(2)}(t) = y_{2,1}^{(2)}(t)$ when transmitting ``1". Therefore, WD$_2$ harvests the following average signal power during phase II,
\begin{equation}
\label{e22}
\begin{aligned}
P_2^{(2)}& =\eta\beta \left[p_0E\big[|y_{2,0}^{(2)}(t)|^2\big] +(1-p_0)E\big[|y_{2,1}^{(2)}(t)|^2\big] \right]\\
&= \eta\beta P_1\left[p_0h_2+(1-p_0)|\alpha_2+ \mu \alpha_1 \alpha_{12}|^2\right],
\end{aligned}
\end{equation}
where $p_0$ denotes the probability of transmitting ``0". Without loss of generality, we consider $p_0=0.5$ in the following analysis. Because a large number of i.i.d. random bits are sent during the backscattering stage (e.g., more than several thousand bits in practice), the amount of energy harvested by WD$_2$, denoted by $Q^{(2)}_2$, can be well characterized by the following scaled average harvest energy,
\begin{equation}
E_2^{(2)} =  \omega t_2 P^{(2)}_2
= \frac{1}{2}\omega\eta t_2\beta  P_1(h_2+|\alpha_2+ \mu \alpha_1 \alpha_{12}|^2),
\end{equation}
where  $\omega\in (0,1]$ denotes a power margin parameter to ensure that $Pr[Q^{(2)}_2 \geq E^{(2)}_2] > 1- \sigma$ by the central limit theorem and $\sigma$ is a small parameter. In other words, WD$_2$ can harvest more than $E^{(2)}_2$ with sufficiently high probability, thus we can safely use $E^{(2)}_2$ to represent the energy harvested by the WD$_2$ during phase II in the following. Meanwhile, it is assumed that WD$_1$ keeps its battery level unchanged during this phase, where the small amount of harvested energy is used for powering the backscatter transmit circuit \cite{2013:Vli}.

We denote the sampling rate of backscatter receiver at WD$_2$ as $R_s=NR_b$, i.e., the number of samples in the transmission of a bit information is $N$. We consider using an optimal energy detector to decode the one-bit information, where the bit error rate (BER) is shown in the following lemma.

\underline{\emph{Lemma}} \emph{3.1}: The bit error rate (BER) $\epsilon$ of the optimal energy detector is
 \begin{equation}
 \label{Pb}
\epsilon={\frac{1}{2}}\text{erfc}\left[{\frac{(1-\beta)P_1\mu^2 h_1h_{12}\sqrt{N}}{4\big((1-\beta)N_0+N_s\big)}}\right],
\end{equation}
where erfc($\cdot$) is the complementary error function defined as
\begin{equation}
\text{erfc}(x) =\frac{2}{\sqrt{\pi}}\int_{x}^{\infty}e^{-t^2}dt.\\
\end{equation}

\emph{Proof:} Please refer to Appendix 1.

With the optimal energy detector, the backscatter communication is equivalent to a binary symmetric communication channel with a cross error probability $\epsilon$. Thus, we can express the channel capacity as
\begin{equation}
C = 1+\epsilon \log_2\epsilon+(1-\epsilon )\log_2(1-\epsilon).
\end{equation}
Accordingly, the date rate of WD$_1$ transmitting to WD$_2$ is
\begin{equation}
\label{R11}
R_1^{(1)}(\mathbf{t}) = C R_b t_2.
\end{equation}

\emph{Remark \text{1}}: Given a fixed sampling rate $R_s$, a larger $R_b$ leads to a smaller $N$, thus higher BER in (\ref{Pb}). Consider an extreme case that $R_b\rightarrow \infty$, we have $N\rightarrow 0$ and $\epsilon \rightarrow 0.5$. Accordingly, the channel capacity $C\rightarrow 0$, resulting a zero data rate $R^{(1)}_1(\mathbf{t})$ in (\ref{R11}). Therefore, a higher backscatter rate $R_b$ does not directly translate to a higher effective data rate due to the higher decoding error probability.

\subsection{Phase \uppercase\expandafter{\romannumeral3}: Active Information Transmission}
 After the backscattering communication phase, WD$_1$ continues to transmit information in active communication mode in Phase III, which exhausts the energy harvested in phase I. Accordingly, the transmit power of WD$_1$ is
 \begin{equation}
 \label{P3}
P_3=E_1^{(1)}/{t_3}={\eta}{P_1}{h_1}{t_1}/{t_3}.
\end{equation}
We denote the complex base-band signal transmitted by WD$_1$ in Phase III as $x_3(t)$ with $E[|x_3(t)|^2] =1$, such that WD$_2$ and the HAP respectively receive
\begin{equation}
\label{y203}
y_2^{(3)}(t)=\alpha_{12}\sqrt{P_3}{x_3(t)}+n_2^{(3)}(t),
\end{equation}
\begin{equation}
y_0^{(3)}(t)=\alpha_1\sqrt{P_3}{x_3(t)}+n_0^{(3)}(t),
\end{equation}
where $n_2^{3}(t)$ and $n_0^{(3)}(t)$ denote the independent Gaussian receiver noises both with power $N_0$. Thus, the achievable rates from WD$_1$ to WD$_2$ and WD$_1$ to the HAP in phase III are
\begin{equation}
\label{Ry12}
R_1^{(2)}(\mathbf{t},\mathbf{P})=\frac{t_3}{T}B\log_{2}\left(1+\frac{P_3h_{12}}{N_0}\right),
\end{equation}
\begin{equation}
\label{Ry13}
R_1^{(3)}(\mathbf{t},\mathbf{P})=\frac{t_3}{T}B\log_{2}\left(1+\frac{P_3h_1}{N_0}\right),
\end{equation}
where $B$ denotes the system bandwidth and it is assumed without loss of generality that $T=1$, such that $T$ is not present in (\ref{Ry12}) and (\ref{Ry13}) as well as the data rate expressions in the remainder of this paper.
\subsection{Phase \uppercase\expandafter{\romannumeral4}: Information Relaying}
In the last phase of duration $t_4$, WD$_2$ first relays WD$_1$'s message with transmit power $P_{41}$ for $t_{41}$ amount of time, then transmits its own message to the HAP with power $P_{42}$ and duration $t_{42}$. Thus, the total energy consumption on WD$_2$ is restricted by the total energy harvested in the first two phases, i.e.,
\begin{equation}
\label{con}
t_{41}P_{41}+t_{42}P_{42}\leq E_2^{(1)}+E_2^{(2)}.
\end{equation}

We denote the time and power allocations as $\mathbf{t}=[t_1,t_2,t_3,t_{41},t_{42}]$ and $\mathbf{P}=[P_1,P_3,P_{41},P_{42}]$, respectively. Then, the transmission rate of WD$_2$ relaying WD$_1$'s information to the HAP is
\begin{equation}
\label{Ry14}
R_1^{(4)}(\mathbf{t},\mathbf{P})=t_{41}B\log_{2}\left(1+\frac{P_{41}h_2}{N_0}\right).
\end{equation}

Note that the HAP can jointly decode WD$_1$'s active information transmission in the $3$-rd and $4$-th phases. Therefore, the achievable rate of WD$_1$ in the time period of duration $T=1$ is\cite{2014:Ju2}
\begin{equation}\label{11}
\begin{aligned}
 R_1(\mathbf{t},\mathbf{P}) = \min [ &R_1^{(1)}(\mathbf{t})+R_1^{(2)}(\mathbf{t},\mathbf{P}),\\&R_1^{(3)}(\mathbf{t},\mathbf{P})+R_1^{(4)}(\mathbf{t},\mathbf{P})],
\end{aligned}
\end{equation}
and WD$_2$'s achievable rate is
\begin{equation}
\label{22}
R_2(\mathbf{t},\mathbf{P})=t_{42}B\log_{2}\left(1+\frac{P_{42}h_2}{N_0}\right).
\end{equation}

\emph{Remark \text{2}}: The proposed backscatter-assisted relaying reduces to the conventional active two-user cooperation in WPCN (e.g., in {\cite{2014:Ju2}}) when phase II is eliminated (i.e., $t_2 =0$). Further, if we set $t_2 = t_{41} = 0$, the proposed method reduces to the case that the two users transmit each independent message to the HAP without cooperation {\cite{2014:Ju1}}. In other words, they are both special cases of ours.

\subsection{Problem Formulation}
In this paper, we jointly optimize the time allocation $\mathbf{t}$ and power allocation $\mathbf{P}$ on wireless
energy and information transmissions to maximize the minimum (max-min) throughput of the two users. The optimal solution is often referred to as the $\emph{common throughput}$, which directly reflects the user fairness in the network. Mathematically, the max-min throughput optimization problem is
\begin{equation}
\label{1}
   \begin{aligned}
    (\rm{P1}): & \max_{\mathbf{t},\mathbf{P}} & &  \min(R_1(\mathbf{t},\mathbf{P}),R_2(\mathbf{t},\mathbf{P}))\\
    &\text{s. t.}    & &(\ref{t}) , (\ref{P3}),\ \rm{and}\ (\ref{con}),\\
    & & & t_1,t_2,t_3,t_{41},t_{42}\geq 0,\\
    & & & P_3,P_{41},P_{42}\geq 0.
   \end{aligned}
\end{equation}

In the next section, we propose an effective optimization algorithm to solve (P1). It is worth mentioning that the proposed solution algorithm  can also be extended to solve the weighted sum rate (WSR) maximization problem of the two users, i.e., maximizing $\omega_1R_1(\mathbf{t},\mathbf{P})+\omega_2R_2(\mathbf{t},\mathbf{P})$ given two fixed positive weighting parameters $\omega_1$ and $\omega_2$ ($\omega_1+\omega_2=1$). The detailed solution methods are omitted for brevity, while the WSR performance will be demonstrated in Simulations when discussing the achievable rate region.

\section{Optimal Solution to (P1)}
\subsection{Problem Reformulation}
 We observe that problem (P1) is non-convex because of the multiplicative terms in (\ref{con}). By introducing two auxiliary variables $\tau_{41}=t_{41}P_{41}$ and $\tau_{42}=t_{42}P_{42}$,  (P1) is transformed into a convex problem. With $P_3$ in (\ref{P3}), we can express $R_1^{(2)}(\mathbf{t},\mathbf{P}), R_1^{(3)}(\mathbf{t},\mathbf{P})$, and $R_1^{(4)}(\mathbf{t},\mathbf{P})$ in (\ref{Ry12}), (\ref{Ry13}) and (\ref{Ry14}) as functions of $\mathbf{t}$. Meanwhile, $R_1(\mathbf{t},\mathbf{P})$ and $R_2(\mathbf{t},\mathbf{P})$ in (\ref{11}) and (\ref{22}) are reformulated as functions of $\mathbf{t}$ and $\boldsymbol{\tau}=[\tau_{41},\tau_{42}]$, i.e.,
\begin{equation}
\label{Ry122}
R_1^{(2)}(\mathbf{t})=t_3B\log_{2}\left(1+{\rho_1^{(2)}}{\frac{t_1}{t_3}}\right),
\end{equation}
\begin{equation}
\label{Ry133}
R_1^{(3)}(\mathbf{t})=t_3B\log_{2}\left(1+{\rho_1^{(3)}}{\frac{t_1}{t_3}}\right),
\end{equation}
\begin{equation}
\label{Ry144}
R_1^{(4)}(\mathbf{t},\boldsymbol{\tau})=t_{41}B\log_{2}\left(1+{\rho_2}{\frac{\tau_{41}}{t_{41}}}\right),
\end{equation}
\begin{equation}\label{111}
\begin{aligned}
 R_1(\mathbf{t},\boldsymbol{\tau}) = \min [&R_1^{(1)}(\mathbf{t})+R_1^{(2)}(\mathbf{t}),\\ & R_1^{(3)}(\mathbf{t})+R_1^{(4)}(\mathbf{t},\boldsymbol{\tau})],
 \end{aligned}
\end{equation}
\begin{equation}
\label{222}
R_2(\mathbf{t},\boldsymbol{\tau})=t_{42}B\log_{2}\left(1+{\rho_2}{\frac{\tau_{42}}{t_{42}}}\right),
\end{equation}
where $\rho_1^{(2)}=h_1h_{12}{\frac{\eta P_1}{N_0}}$, $\rho_1^{(3)}={h_1^2{\frac{\eta P_1}{N_0}}}$, $\rho_2=\frac{h_2}{N_0}$ are constant parameters.

Consequently, we introduce another auxiliary variable $\bar{R}$ and transform problem (P1) into the following equivalent problem (P2):
\begin{equation}
\label{2}
 \begin{aligned}
    (\rm{P2}): & \max_{\overline{R},\mathbf{t},\boldsymbol{\tau}} & &  \overline{R} \\
    &\text{s. t.}   & & t_1,t_2,t_3,t_{41},t_{42}\geq 0,\\
     & & & \tau_{41},\tau_{42}\geq 0,\\
    & & & t_0+t_1+t_2+t_3+t_{41}+t_{42}\leq 1, \\
    & & & \tau_{41}+\tau_{42}\leq E_2^{(1)}+E_2^{(2)},\\
    & & & \overline{R}\leq R_1^{(1)}(\mathbf{t})+R_1^{(2)}(\mathbf{t}),\\
    & & & \overline{R}\leq R_1^{(3)}(\mathbf{t})+R_1^{(4)}(\mathbf{t},\boldsymbol{\tau}),\\
    & & & \overline{R}\leq R_2(\mathbf{t},\boldsymbol{\tau}).
    & & &
 \end{aligned}
\end{equation}

The following lemma shows that (P2) is a convex optimization problem. Therefore, it can be solved using classic convex optimization algorithms (such as interior point method \cite{2004:Boyd}). When the optimal $\boldsymbol{\tau}^*$ and $\mathbf{t}^*$ are obtained, the optimal power allocation $\mathbf P^*$ in (P1) can be easily obtained as $P_{41}^* = \tau_{41}^*/t_{41}^*$ and  $P_{42}^* = \tau_{42}^*/t_{42}^*$.

\underline{\emph{Lemma}} \emph{4.1}:
$R_1^{(2)}(\mathbf{t}),R_1^{(3)}(\mathbf{t}),R_1^{(4)}(\mathbf{t},\boldsymbol{\tau})$ and $R_2(\mathbf{t},\boldsymbol{\tau})$ are concave functions.

\emph{Proof:} Please refer to Appendix 2.
\subsection{Alternative Solution Method}
To obtain some insights on the optimal solution structure and further reduce the complexity of general convex optimization algorithms for solving (P2), we derive in this subsection an alternative method to solve (P2). Specifically, a partial Lagrangian of (P2) is given by
\begin{equation}
\label{lgr}
\begin{aligned}
\mathcal{L}(\overline{R},\mathbf{t},\boldsymbol{\tau,\lambda})=\overline{R}&-\lambda_1(t_0+t_1+t_2+t_3+t_{41}+t_{42}-1)\\ &-\lambda_2(\tau_{41}+\tau_{42}-E_2^{(1)}-E_2^{(2)})\\
&-\lambda_3\left(\overline{R}-R_1^{(1)}(\mathbf{t})-R_1^{(2)}(\mathbf{t})\right)\\&-\lambda_4\left(\overline{R}-R_1^{(3)}(\mathbf{t})-R_1^{(4)}(\mathbf{t},\boldsymbol{\tau})\right)\\
&-\lambda_5\left(\overline{R}-R_2(\mathbf{t},\boldsymbol{\tau})\right),
\end{aligned}
\end{equation}
where $\boldsymbol{\lambda}=[\lambda_1,\lambda_2,\lambda_3,\lambda_4,\lambda_5]$ denotes the Lagrange multipliers associated with the corresponding constraints in (\ref{2}). We can express the dual function of (P2) as
\begin{equation}
\label{dde}
 \begin{aligned}
    d(\boldsymbol\lambda)= & ~\max_{\overline{R}, \mathbf t,\boldsymbol\tau} & &  \mathcal{L}(\overline{R},\mathbf{t},\boldsymbol{\tau,\lambda}) \\
    &~~\text{s. t.}   & & \overline{R},\mathbf{t},\boldsymbol{\tau}\geq 0,
\end{aligned}
\end{equation}
and the dual problem is
\begin{equation}
\label{dual}
 \begin{aligned}
    (\rm{P3}):& ~\min_{\boldsymbol\lambda} & &  d(\boldsymbol{\lambda}) \\
    &~~\text{s. t.}   & & \boldsymbol{\lambda}\geq 0.
\end{aligned}
\end{equation}
The optimal solution $\mathbf{t}^*$ can be obtained if the optimal dual solution $\boldsymbol{\lambda}^*$ is found by solving the dual problem of (P2). We first investigate the optimal solution of the dual function in (\ref{dde}) given a set of dual variables. The first-order necessary conditions for maximizing the dual function are
\begin{equation}
\label{d1}
\frac{\partial\mathcal{L}}{\partial \overline{R}}=1-\lambda_3-\lambda_4-\lambda_5=0,
\end{equation}
\begin{equation}
\label{d2}
\frac{\partial\mathcal{L}}{\partial t_1}=-\lambda_1+\eta P_1 h_2\lambda_2+\frac{B}{\ln2}\left(\frac{\lambda_3\rho_1^{(2)}}{1+\rho_1^{(2)}\frac{t_1}{t_3}}+
\frac{\lambda_4\rho_1^{(3)}}{1+\rho_1^{(3)}\frac{t_1}{t_3}}\right)=0,
\end{equation}
\begin{equation}
\label{d3}
\frac{\partial\mathcal{L}}{\partial t_2}=-\lambda_1+\frac{1}{2}\omega\eta\beta  P_1(h_2+|\alpha_2+ \mu \alpha_1 \alpha_{12}|^2)\lambda_2+CR_b\lambda_3=0,
\end{equation}
\begin{equation}
\label{d5}
\frac{\partial\mathcal{L}}{\partial t_{41}}=-\lambda_1+\frac{\lambda_4B}{\ln2}\left(\ln\left(1+\rho_2\frac{\tau_{41}}{t_{41}}\right)-\frac{\rho_2\frac{\tau_{41}}{t_{41}}}{1+\rho_2\frac{\tau_{41}}{t_{41}}}\right)=0,
\end{equation}
\begin{equation}
\label{d6}
\frac{\partial\mathcal{L}}{\partial t_{42}}=-\lambda_1+\frac{\lambda_5B}{\ln2}\left(\ln\left(1+\rho_2\frac{\tau_{42}}{t_{42}}\right)-\frac{\rho_2\frac{\tau_{42}}{t_{42}}}{1+\rho_2\frac{\tau_{42}}{t_{42}}}\right)=0.
\end{equation}
From (\ref{d1}) and (\ref{d3}), we see that the dual variables $\boldsymbol\lambda$ must satisfy the two equalities for the dual function to be bounded above. Suppose that (\ref{d1}) and (\ref{d3}) are satisfied, we derive the optimal solution of (\ref{dde}) as follows. By introducing a new variable $z_1=\frac{t_1}{t_3}$, (\ref{d2}) can be expressed in the form of $az_1^2+bz_1+c=0$, where
\begin{equation}
a=(\lambda_1^*-\eta P_1 h_2\lambda_2^*)\rho_1^{(2)}\rho_1^{(3)}\ln2,
\end{equation}
\begin{equation}
b=(\lambda_1^*-\eta P_1 h_2\lambda_2^*)(\rho_1^{(2)}+\rho_1^{(3)})\ln2-(\lambda_3^*+\lambda_4^*)B\rho_1^{(2)}\rho_1^{(3)},
\end{equation}
\begin{equation}
c=(\lambda_1^*-\eta P_1 h_2\lambda_2^*)\ln2-\lambda_3^*B\rho_1^{(2)}-\lambda_4^*B\rho_1^{(3)}.
\end{equation}
Since $t_1^*,t_3^*\geq0$ hold at the optimum, we only select the positive solution to the quadratic equality, where
\begin{equation} \label{t1}
z_1^*=\frac{t_1^*}{t_3^*}=\frac{\sqrt{b^2-4ac}-b}{2a}.
\end{equation}
Similarly, by changing variables as $z_{41}=\rho_2\frac{\tau_{41}}{t_{41}}$ and $z_{42}=\rho_2\frac{\tau_{42}}{t_{42}}$ in (\ref{d5}) and (\ref{d6}), we have the following equations
\begin{equation}\label{f41}
\lambda_4^*B\left(\ln(1+z_{41})-\frac{z_{41}}{1+z_{41}}\right)=\lambda_1^*\ln2,
\end{equation}
\begin{equation}\label{f42}
\lambda_5^*B\left(\ln(1+z_{42})-\frac{z_{42}}{1+z_{42}}\right)=\lambda_1^*\ln2.
\end{equation}
Define $f(z)=\ln(1+z)-\frac{z}{1+z}$, which is a monotonically increasing function when $z\geq0$. Therefore, given the dual variables, we can obtain unique $z_{41}^*$ and $z_{42}^*$ as the solutions of $f(z_{41})=\frac{\lambda_1^*\ln2}{\lambda_4^*}$ and $f(z_{42})=\frac{\lambda_1^*\ln2}{\lambda_5^*}$ in (\ref{f41}) and (\ref{f42}), e.g., using the Newton's method. The following Lemma establishes the relation between $t_{41}$ and $\tau_{41}$ ($t_{42}$ and $\tau_{42}$) at the optimum of (P2).

\underline{\emph{Lemma}} \emph{4.2}: The unique optimal $z_{41}^*$, $z_{42}^*$  are expressed as
\begin{equation}
\label{z41}
z_{41}^*=-\left(W\left(-\frac{1}{\text{exp}(1+\frac{\lambda_1^*}{\lambda_4^*B}\ln2)}\right)\right)^{-1}-1,
\end{equation}
\begin{equation}
\label{z42}
z_{42}^*=-\left(W\left(-\frac{1}{\text{exp}(1+\frac{\lambda_1^*}{\lambda_5^*B}\ln2)}\right)\right)^{-1}-1,
\end{equation}
where $W(x)$ denotes the Lambert-W function, which is the inverse function of $f(z)=z\rm{exp}$$(z)=x$, i.e., $z=W(x)$. Accordingly, the optimal power allocation $P_{41}^*$ and $P_{42}^*$ are
$P_{41}^*=\frac{N_0}{h_2}z_{41}^*,P_{42}^*=\frac{N_0}{h_2}z_{42}^*$.

\emph{Proof:} Please refer to Appendix 3.

With the obtained the optimal $z_{41}^*,z_{42}^*$ from (\ref{z41}) and (\ref{z42}), the optimal $t_{41}^*,t_{42}^*$ and $\tau_{41}^*,\tau_{42}^*$  satisfy
\begin{equation}
\label{t41}
\frac{\tau_{41}^*}{t_{41}^*}=\frac{z_{41}^*}{\rho_2},
\end{equation}
\begin{equation}
\label{t42}
\frac{\tau_{42}^*}{t_{42}^*}=\frac{z_{42}^*}{\rho_2}.
\end{equation}

\emph{Remark \text{3}}: It can be easily verified from (\ref{z41}) and (\ref{z42}) that $\lambda_1,\lambda_4,\lambda_5>0$ must hold, which indicates the corresponding constraints of (P2) are active. Using (\ref{z41}) as an example, as $W(x)\in(-1,0)$ when $x\in(-1/e,0)$, if $\lambda_1^*=0$, the optimal solution $\tau_{41}^* = 0$, and if $\lambda_4^*=0$, the optimal $\tau_{41}^* = \infty$. Both cases obviously will not hold at the optimum, therefore $\lambda_1^*=\lambda_4^*=0$. Similar argument also leads to the result that $\lambda^*_5 =0$.

Then, the optimal solution to dual function (\ref{dde}) can be obtained as follows. Notice that any solution $\{\mathbf t,\boldsymbol\tau\}$ satisfying (\ref{t1}), (\ref{t41}) and (\ref{t42}) is optimal to problem (\ref{dde}), thus there are infinite number of equally optimal solutions. We therefore only need to find one particular solution that satisfies the three equalities. For example, we can easily find a set of $\{\mathbf t,\boldsymbol\tau\}$ that satisfies the total time constraint (\ref{t}) in addition to (\ref{t1}), (\ref{t41}) and (\ref{t42}). Then, we substitute the optimal $\mathbf t^*$ to (\ref{111}) and (\ref{222}) to compute $\overline R = \min [R_1(\mathbf{t}^*),R_2(\mathbf{t}^*)] $. This will lead a set of optimal solutions $\{t^*,\boldsymbol\tau^*,\bar{R}^* \} $ of (\ref{dde}).

After solving the dual function, We update the dual variables $\boldsymbol\lambda$ by using the projected sub-gradient method. By substituting the obtained $\{t^*,\boldsymbol\tau^*,\bar{R}^* \} $ to the corresponding terms, we obtain the sub-gradient of the dual variables in $d\boldsymbol{(\lambda)}$, denoted $\boldsymbol\upsilon=[\upsilon_1,\upsilon_2,\upsilon_3,\upsilon_4,\upsilon_5]$ as
\begin{equation}
\label{up1}
\upsilon_1=t_0+t_1+t_2+t_3+t_{41}+t_{42}-1,
\end{equation}
\begin{equation}
\label{up2}
\upsilon_2=\tau_{41}^*+\tau_{42}^*-E_2^{(1)}-E_2^{(2)},
\end{equation}
\begin{equation}
\label{up3}
\upsilon_3=\overline{R}^*-R_1^{(1)}(\mathbf{t}^*)-R_1^{(2)}(\mathbf{t}^*),
\end{equation}
\begin{equation}
\label{up4}
\upsilon_4=\overline{R}^*-R_1^{(3)}(\mathbf{t})-R_1^{(4)}(\mathbf{t}^*,\boldsymbol{\tau}^*),
\end{equation}
\begin{equation}
\label{up5}
\upsilon_5=\overline{R}^*-R_2(\mathbf{t}^*,\boldsymbol{\tau}^*).
\end{equation}
Because the total time constraint in (\ref{t}) is satisfied with equality in the design of dual function optimal solution, the sub-gradient to $\lambda_1$ is always $\upsilon_1=0$. Suppose that an initial feasible $\boldsymbol{\lambda}^{(0)}$ is given, the dual variable $\boldsymbol\lambda$ is updated in the $(k+1)$-th iteration by the following projection to the feasible region of $\boldsymbol\lambda$, denoted by $\mathcal{H}$, i.e.,
\begin{equation}
\label{up}
\begin{matrix}
\boldsymbol\lambda^{(k+1)}=\prod_{\mathcal{H}}(\boldsymbol\lambda^{(k)}-\alpha \boldsymbol\upsilon),
\end{matrix}
\end{equation}
where $\alpha$ is a small learning rate. Specifically, the above projection is calculated from the following convex problem,
\begin{equation}
\begin{aligned}
\begin{matrix}
&\prod_{\mathcal{H}}(\hat{\boldsymbol\lambda})=&\arg \min_{\boldsymbol\lambda}\left\|\boldsymbol\lambda- \hat{\boldsymbol\lambda}\right\|,\\
&\text{s.t}.&(\ref{d1}), (\ref{d3}),\\
&&\lambda_1,\lambda_2,\lambda_3,\lambda_4,\lambda_5\geq 0,
\end{matrix}
\end{aligned}
\end{equation}
which could be easily solved using a bi-section search over the line connecting $\hat{\boldsymbol\lambda}$ and $\boldsymbol{\lambda}^{(0)}$.
\begin{algorithm}
\footnotesize
\SetAlgoLined
 \SetKwData{Left}{left}\SetKwData{This}{this}\SetKwData{Up}{up}
 \SetKwFunction{Union}{Union}\SetKwFunction{FindCompress}{FindCompress}
 \SetKwInOut{Input}{input}\SetKwInOut{Output}{output}
  \textbf{Initialize}: $k \leftarrow 0$, $\varepsilon \leftarrow 0.001$, $\boldsymbol{\lambda}^{(0)} \ge0$ that satisfies (\ref{d1}) and (\ref{d3});\\
  \Repeat{$\left\|\boldsymbol{\lambda}^{(k+1)}-\boldsymbol\lambda^{(k)}\right\|\le \varepsilon$}{
  Calculate $z_1^*$, $z_{41}^*$ and $z_{42}^*$  using (\ref{t1}), (\ref{z41}) and (\ref{z42}) with given $\boldsymbol\lambda^{(k)}$;\\
  Find a $\mathbf{t}^*$ that satisfies (1) and (41);\\
  Calculate $\tau_{41}^*$ and $\tau_{42}^*$ from (\ref{t41}) and (\ref{t42}), respectively ;\\
   Calculate $\overline R = \min [R_1(\mathbf{t}^*),R_2(\mathbf{t}^*)] $;\\
   Calculate the sub-gradient of $\boldsymbol\lambda^{(k)}$ using (\ref{up1})-(\ref{up5});\\
   Update $\boldsymbol\lambda^{(k)}$ to $\boldsymbol{\lambda}^{(k+1)}$  by solving (\ref{up});\\
   $k \leftarrow k+1$;\\
}
   Substitute $\frac{t_1^*}{t_3^*}$, $\frac{\tau_{41}^*}{t_{41}^*}$ and $\frac{\tau_{42}^*}{t_{42}^*}$ to (P2) and solve the linear programming problem ;\\
\textbf{Set} $P_{41}^*=\frac{\tau_{41}^*}{t_{41}^*}$ and $P_{42}^*=\frac{\tau_{42}^*}{t_{42}^*}$;\\
\textbf{Return} $\{\bar R^*,t^*,\boldsymbol\tau^*\}$ as an optimal solution to (P2).
\caption{Proposed optimal solution algorithm to (P2) }
\label{alg1}
\end{algorithm}

 After obtaining the updated dual variables $\boldsymbol\lambda$, we can further update the optimal solution to (P2). Such iteration proceeds until a stopping criterion is met. Notice that the purpose of the algorithm is to obtain the  optimal dual variables $\boldsymbol\lambda^*$, from which we can obtain the optimal $\frac{t_1^*}{t_3^*}$, $\frac{\tau_{41}^*}{t_{41}^*}$ and $\frac{\tau_{42}^*}{t_{42}^*}$. After substituting $\{\frac{t_1^*}{t_3^*},\frac{\tau_{41}^*}{t_{41}^*},\frac{\tau_{42}^*}{t_{42}^*}\}$ into (P2), we transform (P2) into a simple linear programming problem, which can be efficiently solved by the simplex method\cite{2004:Boyd}.  Because (\ref{P3}) is convex, the KKT conditions are sufficient for optimality. Once the optimal solution $\{\mathbf t^*$,
$\boldsymbol\tau^*\}$ are obtained, the optimal power allocation at WD$_2$ is obtained as $P_{41}^*=\frac{\tau_{41}^*}{t_{41}^*}$ and $P_{42}^*=\frac{\tau_{42}^*}{t_{42}^*}$.  The pseudo-code of the optimal solution algorithm to (P2) is summarized in Algorithm 1.

\subsection{Benchmark Methods}
 In this subsection, we select two representative benchmark methods for performance comparison. For both methods, it is assumed that CE occupies the same amount of time $t_0$ as the proposed AB-assisted relaying method.
 \begin{enumerate}
  \item \emph{User cooperation without AB}: This corresponds to the method in \cite{2014:Ju2}. In this case, WD$_1$ does not backscatter during the WET phase, and WD$_2$ relays WD$_1$'s active information transmission to the HAP. We jointly optimize the system time duration and user transmit power allocations to maximize the minimum throughput.
  \item \emph{User cooperation with information exchange}: This corresponds to the method in \cite{2017:MM}. In this case, the two WDs are allowed to share their harvested energy to transmit each other's information. The two cooperating WDs first exchange their independent information with each other as to form a virtual antenna array and then transmit jointly to the HAP. We implement the cooperation scheme and maximize the common throughput by optimizing the transmit time allocation on wireless energy and information transmissions. The detailed expressions are omitted here due to the page limit.
  \item \emph{Independent transmission without cooperation}: The non-cooperation method follows the harvest-then-transmit protocol in \cite{2014:Ju1}. Specifically, WD$_1$ and WD$_2$ first harvest energy from the HAP and then transmit independently their information to the HAP, the achievable rates of WD$_1$ and WD$_2$ are
      \begin{equation}
   \label{R1NO}
   R_1(\mathbf{t})=\frac{t_2}{T}B\log_{2}\left(1+\frac{\eta t_1 P_1 h_1^2}{t_2 N_0}\right),
   \end{equation}
   \begin{equation}
   \label{R2no}
   R_2(\mathbf{t})=\frac{t_3}{T}B\log_{2}\left(1+\frac{\eta t_1 P_1h_2^2}{t_3 N_0}\right).
   \end{equation}

   Thus, the corresponding max-min throughput optimization problem is
   \begin{equation}
   \label{3}
   \begin{aligned}
     & \max_{t_1,t_2,t_3} & &  \min(R_1(\mathbf{t}),R_2(\mathbf{t})) \\
    &\text{s. t.}    & & t_0+t_1+t_2+t_3\leq 1, \\
    & & & t_1,t_2,t_3\geq 0.
   \end{aligned}
   \end{equation}
 \end{enumerate}
\section{Simulation Results}

\begin{table}[h]
\def\tablename{\small{Table}}
\caption{System Parameters}
\label{tab:system parameters}
\centering
\begin{tabular}{|c|c|c|}
\hline
\footnotesize{\textbf{Parameter}} & \footnotesize{\textbf{Description}} & \footnotesize{\textbf{Value}}\\
\hline
$P_1$ & \footnotesize{Transmission power of HAP} & $1$ W \\
\hline
$\eta$ & \footnotesize{Energy harvesting efficiency} & $0.6$ \\
\hline
$N_0$ & \footnotesize{Noise power at receiver antenna} & $10^{-12}$ W \\
\hline
$N_s$&\footnotesize{Noise power at ID circuit}&$10^{-12}$ W \\
\hline
$f_c$ & \footnotesize{Carrier frequency} & $915$ MHz \\
\hline
$\lambda$ & \footnotesize{Path-loss factor} & $2.5$  \\
\hline
$G_A $ & \footnotesize{ Antenna power gain} & $2$ dB \\
\hline
$R_s $ & \footnotesize{ Sampling rate} & $2$ MHz \\
\hline
$B$ & \footnotesize{System bandwidth} & $100$ kHz \\
\hline
$\omega$ & \footnotesize{Power margin} &$0.8$ \\
\hline
$t_0$ & \footnotesize{Channel estimation time} & $0.05$ \\
\hline
$\mu$ & \footnotesize{Backscatter reflection coefficient} & $0.8$  \\
\hline
\end{tabular}
\end{table}
In this section, we provide simulation results to evaluate the performance of the proposed backscatter-assisted cooperation scheme.
In all simulations, we use the parameters of Powercast TX91501-1W transmitter with $P=1$ W as the energy transmitter at the HAP, and P2110 Powerharvester as the energy receiver at each WD with $\eta = 0.6$ energy harvesting efficiency. 
Unless otherwise stated, the parameters used in the simulations are listed in Table I, which correspond to a typical outdoor wireless powered sensor network similar to the setups in \cite{2018:Bi} and \cite{2014:Ju2}.
 In addition, we denote $h_i=G_A(\frac{3\times10^8}{4\pi d_i f_c})^\lambda$ as the channel gain, where $d_1$ and $d_2$ denote HAP-to-WD$_1$ distance and HAP-to-WD$_2$ distance, and $d_{12}$ denotes the distance between the two WDs.  

\begin{figure}
  \centering
   \begin{center}
      \includegraphics[width=0.5\textwidth]{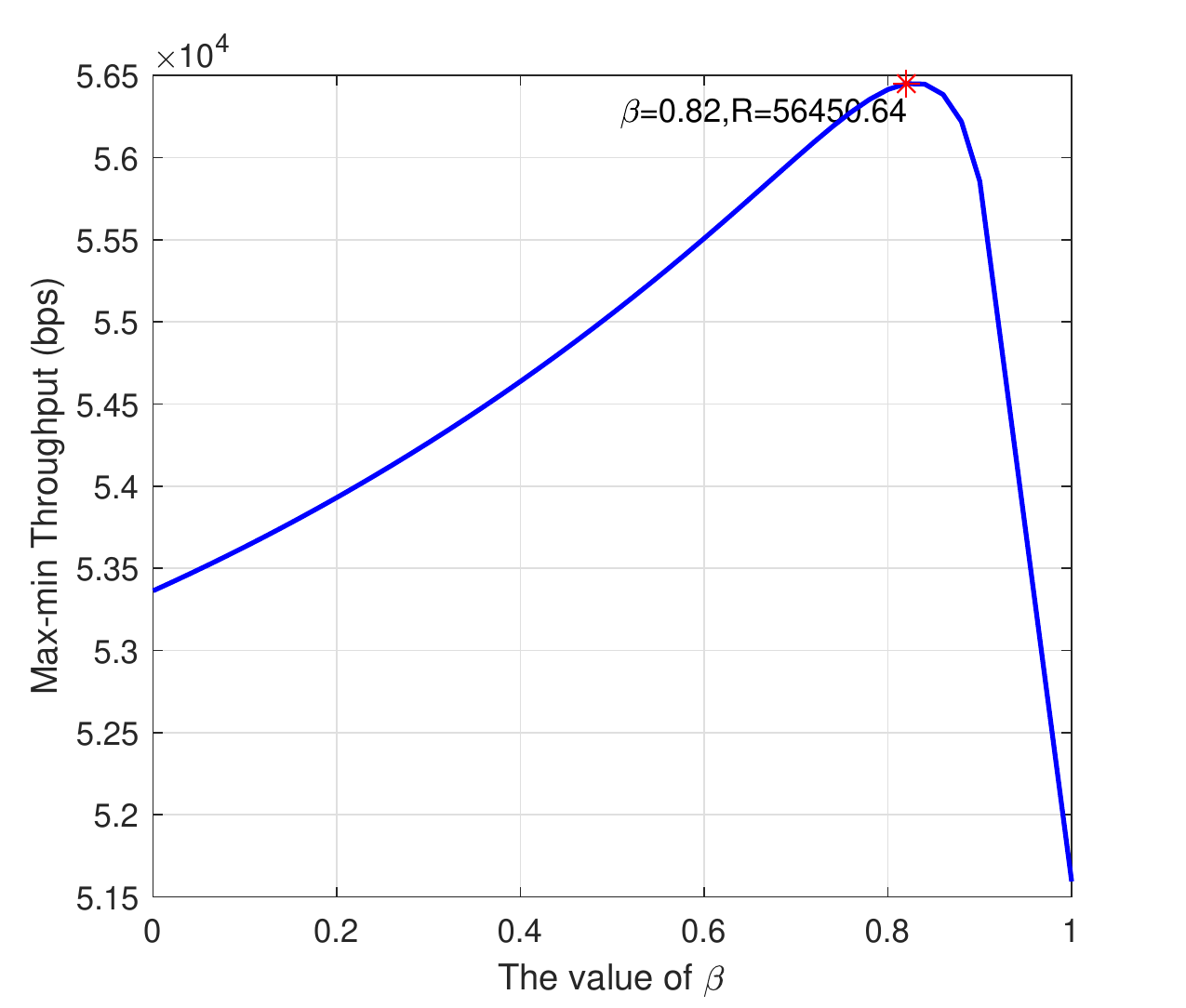}
   \end{center}
  \caption{The max-min throughput versus power splitting factor $\beta$.}
  \label{Fig.4}
\end{figure}

\begin{figure}
  \centering
   \begin{center}
      \includegraphics[width=0.5\textwidth]{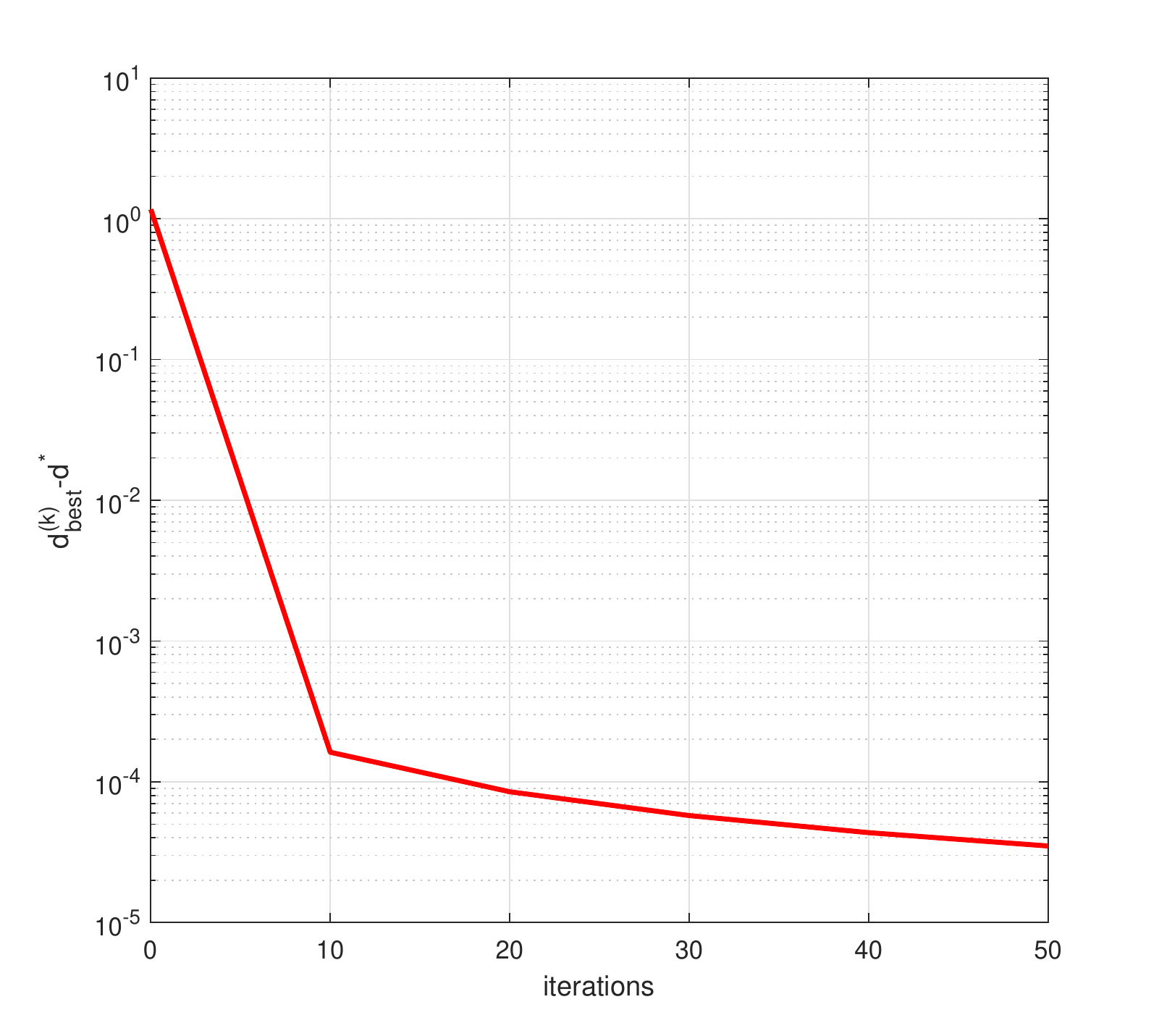}
   \end{center}
  \caption{The value of $d_{best}^{(k)}-d^*$ versus number of iterations $k$.}
  \label{Fig.4.5}
\end{figure}

 We first show in Fig.~\ref{Fig.4} the impact of power splitting factor $\beta$ to the throughput performance. The backscatter rate is set as $R_b$ = 30 kbps. Notice that the backscatter transmission rate depends on the hardware configuration of the wireless devices. Here, we set $h_1 = 1.21 \times 10^{-6}$, $h_2 =3.93\times 10^{-6}$ and $h_{12} = 6.87 \times 10^{-6}$, and change the value of $\beta$ from 0 to 1. Each point in the plot is the optimal throughput performance by solving (P2). It is observed that the minimum transmission rate of two users first increases when $\beta$ increases from 0 and reaches the maximum around 0.8. This is because a larger $\beta$, and thus a larger amount of harvested energy by the energy-constrained WD$_2$ can increase the data transmission rate in the relaying stage. However, as we further increase $\beta$'s value, the throughput performance decreases, this is because the transmission rate from WD$_1$ to WD$_2$ becomes the performance bottleneck due to the reduced SNR at the ID circuit. In general, the optimal value of $\beta$ is related to a number of factors, e.g., device placement and AB communication rate $R_b$, which is not the main focus of this paper. For simplicity of exposition, we assume a fixed  $\beta=0.8$ in the following simulations.

In Fig.~\ref{Fig.4.5}, we plot the convergence performance of the proposed algorithm. Here, we set $h_1 = 1.21 \times 10^{-6}$, $h_2 =3.93\times 10^{-6}$, $h_{12} = 1.41\times 10^{-5}$ and a diminishing step size $\alpha=0.1/k$. It can be seen that the optimality gap decreases quickly to a satisfactory precision (around $10^{-4}$) in less than 10 iterations. Overall, the results show that the proposed primal-dual method has fast convergence property and the overall complexity is low.

\begin{figure}
  \centering
   \begin{center}
      \includegraphics[width=0.5\textwidth]{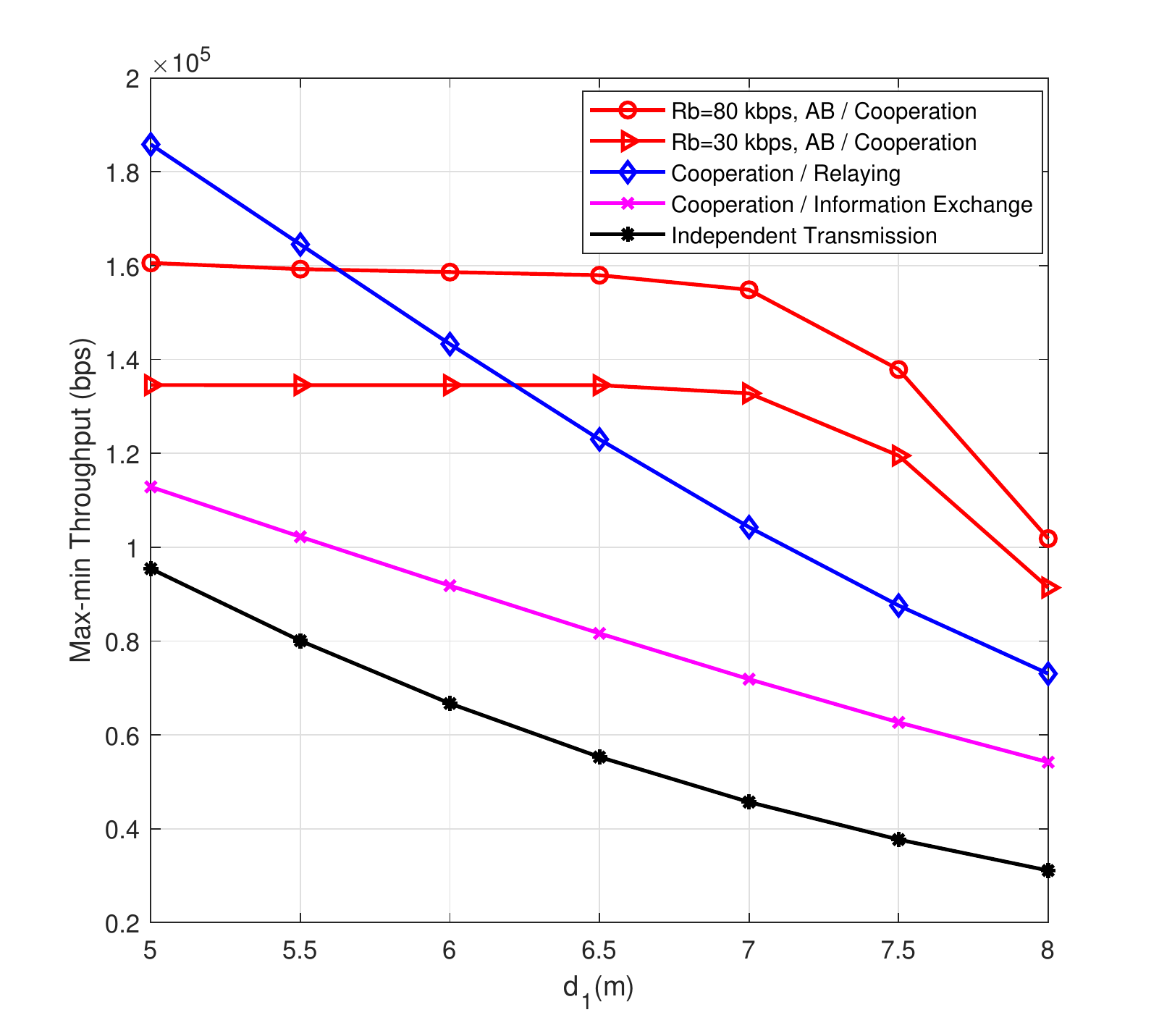}
   \end{center}
  \caption{The max-min throughput performance when the inter-user channel $h_{12}$ varies. Here, we keep $d_2 = 2.5$ meters and vary $d_1$.}
  \label{Fig.5}
\end{figure}

\begin{figure}[h]
  \centering
   \begin{center}
      \includegraphics[width=0.5\textwidth]{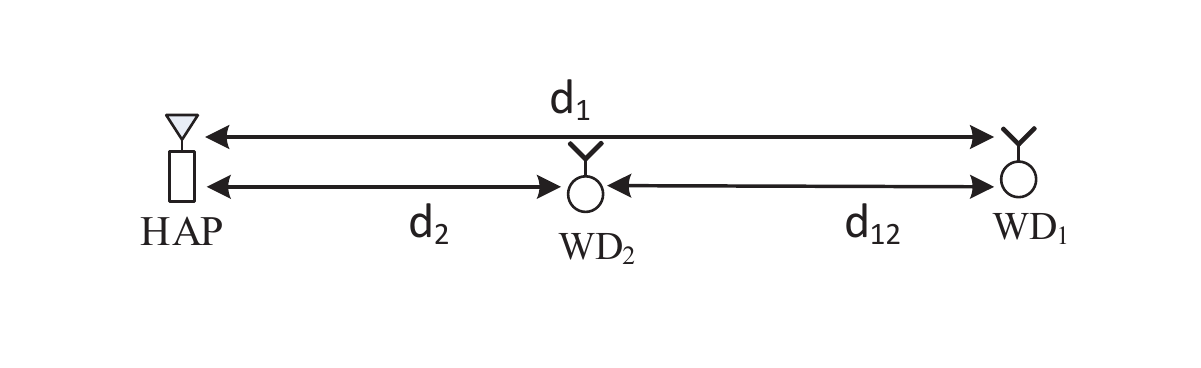}
   \end{center}
  \caption{The line placement model of simulation setup.}
  \label{Fig.6}
\end{figure}

Moreover, in Fig.~\ref{Fig.5}, we numerically show the optimal throughput performance versus the inter-user channel $h_{12}$ for all transmission methods. Besides, we consider the placement model of the network system in Fig.~\ref{Fig.6}, where all the devices are placed on a straight line in which the helping relay user WD$_2$ is in the middle with $d_{12}=d_1-d_2$. Here, we fix $d_2=2.5$ meters and vary $d_1$ from 5 to 8 meters. We consider two different backscatter rates $R_b=30$ kbps, 80 kbps. Obviously, we can see that the max-min throughput decreases when $d_1$ increases for all the methods, because the channel between the two WDs ($h_{12}$) is getting worse when $d_1$ increases. We notice that the proposed backscatter-assisted cooperation method and relaying cooperation method always produce better performance than the cooperation with information exchange method. This is because the information exchange between two users costs significant amount of time and energy. In addition, for the two better-performing cooperation methods, when $R_b = 80$ kbps, we can observe the evident advantage of the proposed backscatter-assisted cooperation method when $d_1 > 5.6$ meters. We can also observe the similar result when $R_b = 30$ kbps, where the performance of the relaying cooperation is worse than the proposed AB-assisted cooperation when $d_1$ is large. This is because when the far user WD$_1$ moves more away from the HAP, it suffers from more severe attenuation in both energy harvesting and information transmission to WD$_2$. Thus, for the relaying cooperation method without AB, the optimal solution
allocates more time for WD$_1$ to harvest energy and transmit information to WD$_2$. It is observed that AB communication can evidently improve the overall throughput performance by reducing the energy and time consumption of information transmission. The performance gain is especially evident when $d_1$ is large, such that the weaker user WD$_1$ is unable to harvest sufficient energy for efficient active information transmission. However, we also see that the communication performance of the AB-assisted cooperation degrades significantly when the inter-user channel is very weak, e.g., $d_1 >7$ meters. This indicates that the cooperation still requires relatively good inter-user channel that the separation of the two cooperating users cannot be too large.

\begin{figure}
  \centering
   \begin{center}
      \includegraphics[width=0.5\textwidth]{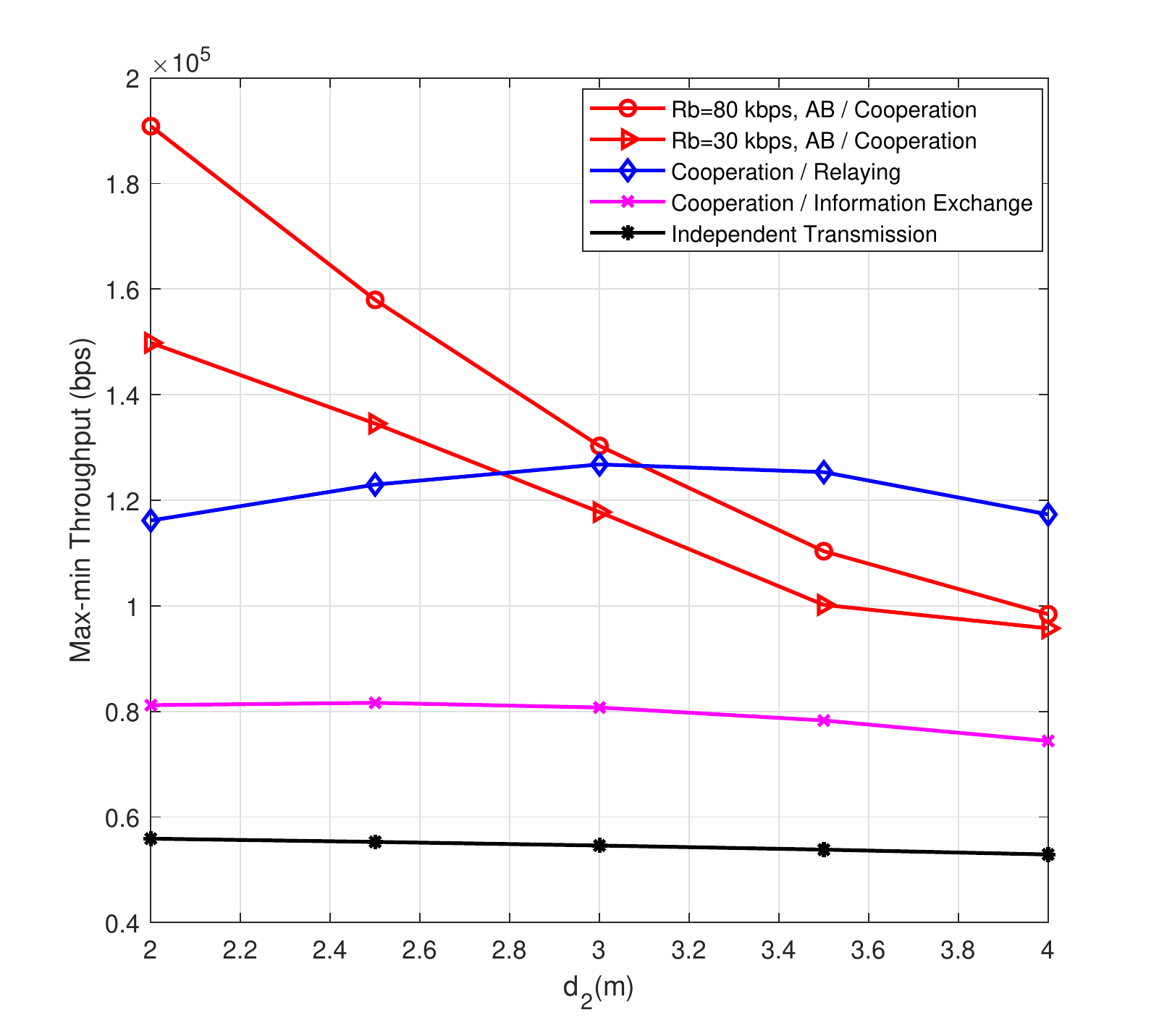}
   \end{center}
  \caption{The max-min throughput performance when the relaying channel $h_2$ varies. Here, we keep $d_1 = 6.5$ meters and vary $d_2$.}
  \label{Fig.7}
\end{figure}

Fig.~\ref{Fig.7} investigates the optimal throughput performance versus the relaying channel $h_{2}$ for all the methods. Here, we still use the line placement model in Fig.~\ref{Fig.6}, where we set $d_1=6.5$ meters and vary $d_2$ from 2 to 4 meters. We first observed that the throughput of the independent transmission method is almost unchanged when $d_2$ increases. That is because no information exchange between the two WDs and its performance mainly depends on  the far user WD$_1$'s weak channel $h_1$. Besides, we notice that the proposed AB-assisted cooperation method and the relaying cooperation method always outperform the cooperation with information exchange method and independent transmission method. For the two better-performing cooperation methods, we also see the proposed backscatter-assisted cooperation method produces better performance when the relay user WD$_2$'s channel $h_2$ is strong ($d_2$ is small). This is because a small $d_2$ results in the weak inter-user channel $h_{12}$. Thus, WD$_1$ needs to consume significant amount of energy on transmitting information actively to the relay user WD$_2$. The considered passive cooperation method can effectively reduce the collaborating overhead and further enhance the transmission performance.

\begin{figure}
  \centering
   \begin{center}
      \includegraphics[width=0.5\textwidth]{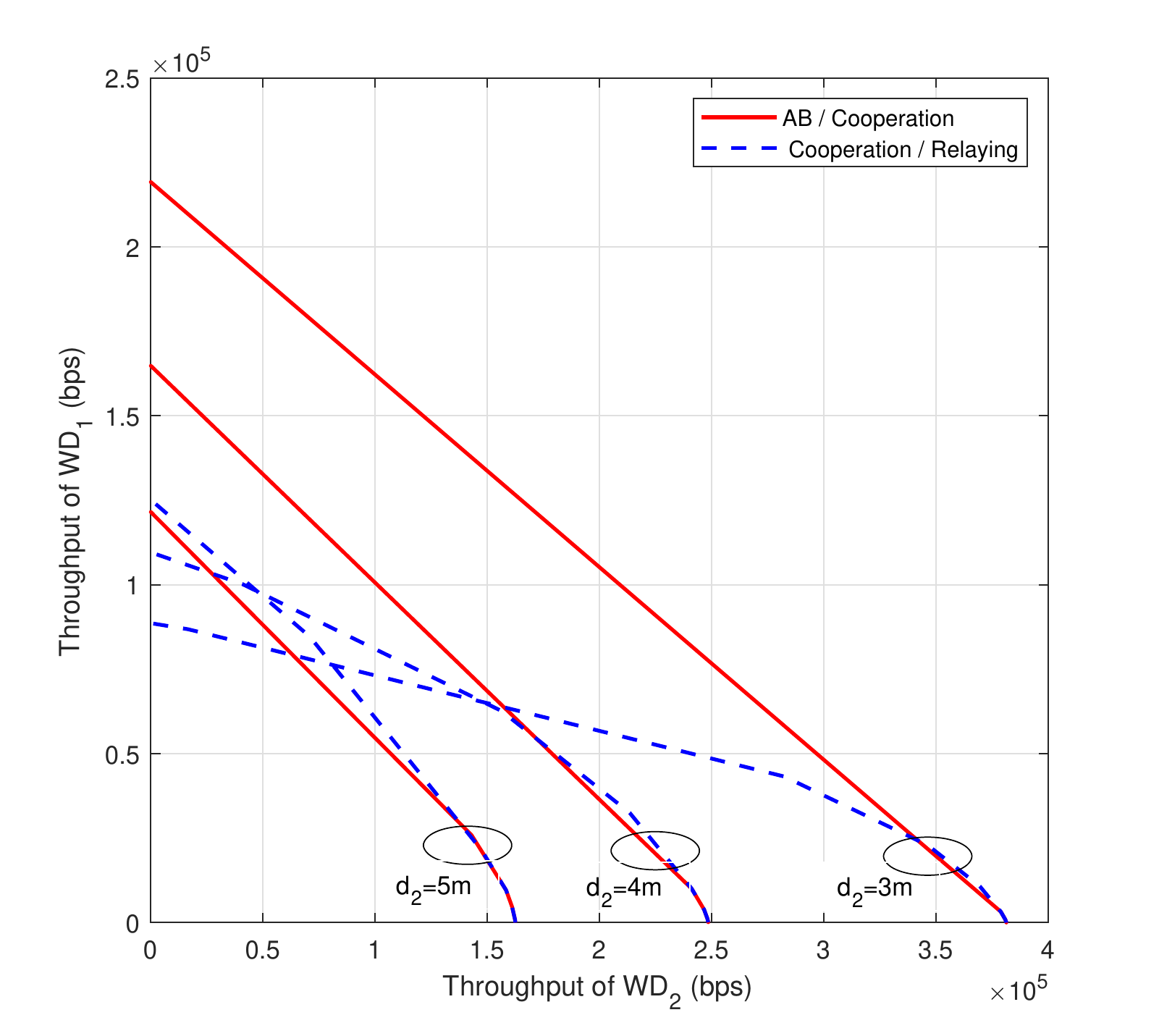}
   \end{center}
  \caption{Throughput region comparison of different methods.}
  \label{Fig.8}
\end{figure}

Fig.~\ref{Fig.8} compares the achievable rate regions of WPCN by solving the weighted sum rate maximization problem when the weighting parameter $\omega_1$ varies from $0$ to $1$. Similarly, we use the line placement model with a fixed $d_1=8$ meters and consider three different distances $d_2=3$ m, 4 m, 5 m. For the two better-performing cooperation methods, the throughput regions of WPCN with the proposed AB-assisted cooperation decreases with increasing $d_2$ due to the inter-user channel $h_{12}$ is getting worse. We observe that the far user's throughput of the considered AB-assisted cooperation is significantly larger than the one without AB communication when $d_2$ is small, and decreases as $d_2$ increases. This is because when the distance between the two WDs is large, it is useful for the AB-cooperation scheme to save the energy needed in the active transmission.
The simulation results in Fig.~\ref{Fig.5}, Fig.~\ref{Fig.7} and Fig.~\ref{Fig.8} demonstrate the advantage of applying backscatter communication to enhance the throughput performance both users when cooperation is considered in WPCN, especially when the channel between the WDs is relatively weak. The advantage mainly comes from the time and energy saving from the simultaneous energy harvesting and passive information exchange enabled by the AB communications.

\section{Conclusions and Future Work}
In this paper, we integrated AB communication and user cooperation in a two-user WPCN. Specifically, the proposed AB-assisted cooperation method achieves simultaneous information transmission in a passive manner by reusing wireless power transfer, which can effectively reduce the transmission time and energy consumption of conventional active communication methods. In addition, we investigated the maximum common throughput optimization problem of the proposed cooperation method, and jointly optimized the time and power allocations of energy-constrained users to obtain the optimal solution, and simulated under extensive network setups to evaluate the performance of the proposed AB cooperation method. By comparing with conventional user cooperation method based on active communication, we showed that the presented backscatter-assisted cooperation method improves the user fairness in WPCN under different practical network setups. Moreover, we also found that the proposed passive cooperation method can significantly save the collaborating overhead (transmission time and energy consumption) and improve the overall throughput performance.

Finally, we conclude the paper with some interesting future working directions. First, it is interesting to consider a more realistic energy harvesting scenario, where  the overhead of setting the value of $\beta$, the energy distribution operation and signal synchronization all affect the energy harvesting performance.  Moreover, although one available way to improve the energy harvesting performance is using two separated architectures, e.g., two separate antennas for energy harvesting and information decoding, however, it may introduce additional production cost to size-constrained IoT devices, we can further study them in our future works. At last, it is also challenging to extend the considered network model to other practical setups, such as multi-user scenario, hybrid  backscatter communication, cluster-based cooperation, and interference channel, etc.

\section*{Appendix 1\\Proof of lemma 3.1}
\emph{Proof:}
Let $B[k]\in \{0,1\}$ denote the information bit transmitted in the backscatter stage,  WD$_2$ receives signal $y_2[i]$ in the proposed user cooperation, ie.,
\begin{equation}
y_2[i]=\alpha_2x_2[i]+B[k]\mu \alpha_1\alpha_{12}x_2[i]+n_2[i],i=1,...,N,
\end{equation}
where $B[k]$ denotes the binary information bits, $n_2 \sim \mathcal {CN}(0 ,N_0)$, and information signal at the WD$_2$ is
\begin{equation}
y[i]=\sqrt{1-\beta}y_2[i]+n_s[i],i=1,...,N,
\end{equation}
where $n_s \sim \mathcal {CN}(0 ,N_s)$, we can express the average  power as
\begin{equation}
E\left[\frac{1}{N} \sum_{i=1}^N \big|y[i]\big|^2\right] =(1-\beta)\big(P_1\big|\alpha_2+B[k]\mu\alpha_1\alpha_{12}\big|^2+N_0\big)+N_s.
\end{equation}

Thus, the corresponding statistical properties are
\begin{equation}
E\left[\sum_{i=1}^N n_s[i]^2\right]=NN_s,\quad
Var\left[\sum_{i=1}^N n_s[i]^2\right]=2NN_s^2.
\end{equation}
 When $N$ is sufficiently large (e.g., $N>10$), we can approximate the test statistic $Z = \frac{1}{N} \sum \limits_{i=1}^N |y[i]|^2$ as a Gaussian random variable by the central limit theorem, i.e.,
\begin{equation}
\label{twodis}
 \begin{aligned}
B[k]=0:Z \sim \mathcal {N}\Big((1-\beta)P_1h_2+(1-\beta)N_0+N_s,\\
\frac{2{((1-\beta)N_0+N_s)}^2}{N}\Big), \\
B[k]=1:Z \sim \mathcal{N}\Big((1-\beta)P_1|\alpha_2+\mu\alpha_1\alpha_{12}|^2+\\
(1-\beta)N_0+N_s,\frac{2{((1-\beta)N_0+N_s)}^2}{N}\Big).
 \end{aligned}
\end{equation}
By defining $Z_1 = Z - (1-\beta)P_1h_2-(1-\beta)N_0-N_s$, we have
\begin{equation}
\label{twodis}
 \begin{aligned}
B[k]=0:&Z_1 \sim \mathcal {N}\Big(0,\frac{2\big((1-\beta)N_0+N_s\big)^2}{N}\Big),\\
B[k]=1:Z_1 \sim \mathcal{N}\Big(&(1-\beta)P_1|\mu^2h_1h_{12}+2\mu\alpha_1\alpha_2\alpha_{12}|,\\
&\frac{2((1-\beta)N_0+N_s)^2}{N}\Big).\\
 \end{aligned}
\end{equation}
 It is assumed without loss of generality that the probability of transmitting ``0"  and ``1"  are equal. Therefore, we can obtain the BER $\epsilon$ as
\begin{equation}
\begin{aligned}
 \epsilon &=\frac{1}{2}\left(P_r\big(\hat{B}(k)=0|B(k)=1\big)+P_r\big(\hat{B}(k)=1|B(k)=0\big)\right)\\
 &=P_r\left((1-\beta)P_1|\frac{1}{2}\mu^2h_1h_{12}+\mu\alpha_1\alpha_2\alpha_{12}|\right)\\&=Q\left({\frac{(1-\beta)P_1\mu^2 h_1h_{12}\sqrt{N}}{2\sqrt{2}\big((1-\beta)N_0+N_s\big)}}\right)\\
 &={\frac{1}{2}}\text{erfc}\left[{\frac{(1-\beta)P_1\mu^2 h_1h_{12}\sqrt{N}}{4\big((1-\beta)N_0+N_s\big)}}\right],
\end{aligned}
\end{equation}
where $Q(\cdot)$ is the Gaussian $Q$-function defined as
\begin{equation}
\begin{aligned}
Q(x) &=\frac{1}{\sqrt{2\pi}}\int_{x}^{\infty}\text{exp}(-\frac{t^2}{2})dt.\\
\end{aligned}
\end{equation}

\section*{Appendix 2\\Proof of lemma 4.1}
\emph{Proof:}
The Hessian of $R_2(\mathbf{t},\boldsymbol{\tau})$ in (\ref{222}) is
\begin{equation}
\label{ll}
\bigtriangledown^2 R_2(t_{42},\tau_{42})=[d_{i,j}],i,j\in[1,2],
\end{equation}
where $d_{i,j}$ can be given by
\begin{equation}
\label{lll}
  d_{i,j}=\left\{
   \begin{aligned}
   -{\frac{{\rho_2^2}{\tau_{42}^2}B}{{t_{42}^{3}}{(1+\rho_2{\frac{\tau_{42}}{t_{42}}})^{2}}\ln2}},      &      & {i=j=1}\\
   {\frac{{\rho_2^2}{\tau_{42}}B}{{t_{42}^{2}}{(1+\rho_2{\frac{\tau_{42}}{t_{42}}})^{2}}\ln2}},     &      & {i\neq j}\\
   -{\frac{{\rho_2^2}B}{{t_{42}}{(1+\rho_2{\frac{\tau_{42}}{t_{42}}})^{2}}\ln2}} ,    &      & {i=j=2}
   \end{aligned}
   \right.
  \end{equation}

Given an arbitrary real vector $\mathbf{v}=[\nu_1,\nu_2]^T$ , we can further obtain from (\ref{ll}) and (\ref{lll}) as
\begin{equation}
\mathbf{v}^T\bigtriangledown^2R_2(t_{42},\tau_{42})\mathbf{v}=-{\frac{{\rho_2^2}B}{{t_{42}}{(1+\rho_2{\frac{\tau_{42}}{t_{42}}})^{2}\ln2}}}{(\frac{\tau_{42}}{t_{42}}\nu_1-\nu_2)^2} \leq 0,
\end{equation}
i.e., $\bigtriangledown^2R_2(t_{42},\tau_{42})$ is a negative semi-definite matrix. Therefore, $R_2(t_{42},\tau_{42})$ is a jointly concave function of both $t_{42}$ and $\tau_{42}$. The proof of $R_1^{(2)}(\mathbf{t}),R_1^{(3)}(\mathbf{t})$ and $R_1^{(4)}(\mathbf{t},\boldsymbol{\tau})$ are all the same as $R_2(\mathbf{t},\boldsymbol{\tau})$.

From Lemma 4.1, we can see that the objective function and the last three constraint conditions of problem (\ref{2}) satisfy the properties of concave function. Furthermore, the constraints from the first four formulas of problem (\ref{2}) are both affine. Thus, problem (P2) is proved to be a convex optimization problem.

\section*{Appendix 3\\Proof of lemma 4.2}
\emph{Proof:}
By solving $f(z_{41})=\frac{\lambda_1^*}{\lambda_4^*B}\ln2$ and $f(z_{42})=\frac{\lambda_1^*}{\lambda_5^*B}\ln2$, where $f(z)=\ln(1+z)-\frac{z}{1+z}$, we have
\begin{equation}
\begin{aligned}
\ln(1+z_{41})+\frac{1}{1+z_{41}}=\frac{\lambda_1^*}{\lambda_4^*B}\ln2+1,\\
\ln(1+z_{42})+\frac{1}{1+z_{42}}=\frac{\lambda_1^*}{\lambda_5^*B}\ln2+1.
\end{aligned}
\end{equation}
After performing the exponential operations at both sides of the above two equations, we obtain
\begin{equation}
\label{42z}
\begin{aligned}
(1+z_{41})\exp\left(\frac{1}{1+z_{41}}\right)=\exp\left(1+\frac{\lambda_1^*}{\lambda_4^*B}\ln2\right),\\
(1+z_{42})\exp\left(\frac{1}{1+z_{42}}\right)=\exp\left(1+\frac{\lambda_1^*}{\lambda_5^*B}\ln2\right).
\end{aligned}
\end{equation}
Consider two positive values $x$ and $z$ that satisfy $\frac{1}{x}\exp(x)=z$, it holds that
\begin{equation}
\label{xe}
-x\exp(-x)=-\frac{1}{z}.
\end{equation}
Thus, we obtain $x=-W(-\frac{1}{z})$, where $W(b)$ is the Lambert-W function and can be obtained
by calculating the inverse function of $f(a)=a\exp(a)=b$, i.e., $a=W(b)$. We can infer from (\ref{42z}) and (\ref{xe}) that $\frac{1}{1+z_{41}}=-W\left(-\frac{1}{\exp(1+\frac{\lambda_1^*}{\lambda_4^*B}\ln2)}\right)$ and  $\frac{1}{1+z_{42}}=-W\left(-\frac{1}{\exp(1+\frac{\lambda_1^*}{\lambda_5^*B}\ln2)}\right)$, as well $z_{41}^*=\rho_2\frac{\tau_{41}^*}{t_{41}^*}=\frac{h_2}{N_0}P_{41}^*$ and  $z_{42}^*=\rho_2\frac{\tau_{42}^*}{t_{42}^*}=\frac{h_2}{N_0}P_{42}^*$. Thus, we can obtain the results in Lemma 4.2 through some simple mathematical derivation.

\bibliographystyle{IEEEtran}
\bibliography{ZY}
%
%
%
%

\end{document}